# Improve the Field Strength by Adding Soft Iron in the Hybrid Permanent magnet

Quanling Peng[1,2], Jianxin Zhou[1], Saike Tian[1], Yingzhe Wang[1]
1. Institute of High Energy Physics, Chinese Academy of sciences, Beijing, 100049, China
2. School of electronics, Nanchang institute of Technology, Jiangxi, 330013, China.

**Abstract**—Permanent magnet has a small and compact structure, is especially suitable for a narrow space. With the aid of soft iron, the magnetic field can be increased much more and the field uniformity can be well controlled. Most Permanent magnets have a symmetry structure; the soft iron can be selected as the float magnet pole to keep its constant scalar potential, and take roles as media to collect the direct flux from the permanent blocks, then release indirect flux in magnet aperture and to the nearby return yoke. This paper presents the magnetic flux method to design and fabricate a hybrid permanent dipole by using axially and radially magnetized permanent blocks. A variable gradient permanent quadrupole and a variable gradient sextupole are designed as the extend design examples . They all consist of two nested hybrid permanent rings, where the iron poles are used to control the field quality, collect the magnetic flux from the outer ring, block the skew quadrupole and high order harmonics.

## 1. Introduction

Electromagnet or superconducting magnets, with field strength varies along with the beam energy, are widely used in modern particle accelerators to bend or focus the particle beams. In some circumstances, where the beam energy is fixed or only a little adjustment, permanent magnet will be a better selection, since it has advantages of small space occupation, no cooling system, and no operation costs, one time investment can maintain a long time operation.

Permanent magnet can be made of all permanent blocks or permanent blocks mixed with soft iron, where the former called as pure permanent magnet, later called as the hybrid permanent magnet. For pure permanent magnet design, K. Halbach had given the design principle in 1980s [1-2]. Each permanent block can be treated as air with the current loops surrounding the magnet or as magnetic charges on its surfaces along the easy axis [3-4]. Field strength in magnet aperture is collected the contribution in each permanent block by its position and easy axis orientation.

An important issue is that field strength from the pure permanent magnet is weak compared with electromagnets or superconducting magnets. According to reference [1], even with the highest residual field strength, maximum field of a dipole that built with pure permanent blocks is less than the remnant field of Br, say about 1.4 T for the highest NdFeB magnetic materials. On the other hand, pure permanent magnet needs permanent blocks with different easy axis orientation, that will bring fabrication difficulties and increase the cost [5-9]. For a hybrid permanent magnet, on the other hand, with the nonlinear material such as iron poles to collect the surrounding magnetic flux and release it into a small compact space, it can reach to a higher magnetic field. Another way to increase the field strength is to add more permanent blocks surround the iron pole to add more magnetic flux. Variable field permanent magnets can be built by adding auxiliary coils around the iron pole or adding an outer concentric permanent ring surround the inner permanent ring [10].

## 2. Principle of flux method

Assume a permanent magnet has an infinite length, the 3D case can be treated in 2D case. In 2D analysis, a homogeneously magnetized permanent block can be treated as an air space with the surface charge density $\sigma^{\pm}=\pm B_r$ on its upper and lower surfaces or a surface current distribution of density $i = H_c$ around the other four surfaces [3-4]. Hear $B_r$ and $H_c$ are the remnant field and the coercively of the permanent block respectively, with $\boldsymbol{B_r}=\mu_0\boldsymbol{H_c}$. For the hard permanent magnetic material, the slope of demagnetized curve in the second quarter is near 1, or $\mu_r=1$, no extra field contribution from the material magnetization.

In 2D non-current space, magnetic field can be expressed as the negative gradient of scalar potential as $\mathbf{B} = -\nabla V$. As shown in Fig. 1, in a hybrid permanent magnet, the permanent block can be modeled by charge sheets at the top and bottom surfaces. The iron pole has large relative permeability, all the surface of the magnet pole keep a constant scalar potential of $V_2=V_0$, the return yoke and mid-plane are all kept in zero scalar potential. Since $V_1$ and $V_2$ are in different scalar potential, magnetic field will be produced between the magnet pole to the mid-plane or to the return yoke, which call as useful field and stray field respectively.

@Manuscript received Jan 2, 2023. Work supported by the accelerator research program of the Chinese Academy of Sciences Grant No. Y5294104TD. Author email address: pengql@ihep.ac.cn.

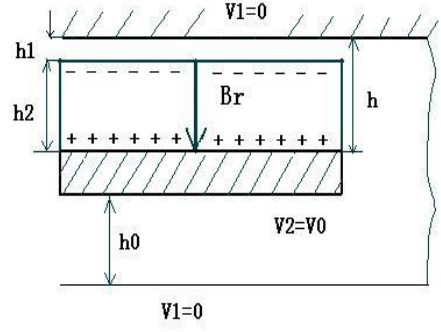

Fig. 1 Different scalar potentials in a quarter of the hybrid permanent dipole, the shade region represent the iron yoke and iron pole.

## 2.1 Direct flux $\phi_d$

Direct flux is magnetic flux coming from the permanent magnet blocks and deposits on the iron pole, it is the source to maintain iron pole in a higher scalar potential. The direct flux to the iron pole is

$$\phi_\sigma = c\sigma D, \quad c = \frac{V(r_i)}{V_0}. \tag{1}$$

$c$ is the fraction of magnetic charge $\sigma$ deposited on the surface of the iron pole, it equals to the scalar potential $V$ at the magnetic charges with respect to the potential $V_0$ at the iron pole, $D$ is the width of the permanent block. Assume the scalar potential changes uniformly in the permanent block, then $c = \frac{h_i}{h}$. In Fig. 1, the direct flux includes the contributions from the positive and the negative charges, which can be expressed as

$$\phi_d = \phi_{\sigma^+} + \phi_{\sigma^-} = D\sigma^+ c_1 + D\sigma^- c_2, \tag{2}$$

here $c_1 = \frac{V(r_{\sigma^+})}{V_0} = 1$, $c_2 = \frac{V(r_{\sigma^-})}{V_0} = \frac{h_1}{h}$. If magnet blocks directly touch on the iron pole and leave some space $h_1$ between the top yoke, eq. (2) can be written as:

$$\phi_d = B_r D \left(1 - \frac{h_1}{h}\right) = B_r D \frac{h_2}{h}. \tag{3}$$

If the permanent block fully occupies the space between the iron pole and top yoke, then the direct flux on the iron pole is:

$$\phi = B_r D. \tag{4}$$

In 3d case, the direct flux deposited on the iron pole is:

$$\Phi = B_r S, \tag{5}$$

S is the surface area of the permanent block.

## 2.2 Indirect flux $\phi_i$

Indirect flux escapes from the iron pole faces to the nearby zero scalar potential areas. As shown in Fig. 2, add permanent magnet between the magnet pole and the side yoke, the scalar potential of the iron pole will rise up to $V_0$, the back yoke, the top yoke and the mid-plane still keeps zero scalar potential, parasite magnetic field will be produced between the different scalar potential areas. Here $\varphi_{i1}$ is the expected field, $\varphi_{i2}$ is the demagnetization field for the permanent magnet, $\varphi_{i3}$ is the nearby leakage field, total indirect flux is $\varphi_i = \varphi_{i1} + \varphi_{i2} + \varphi_{i3}$.

Assume the expected field $B_0$ keeps constant in the magnet gap, with $B_0 = -\frac{\partial V}{\partial y} = -\frac{V_0}{h_0}$, $h_0$ is half-length of the magnet gap. The indirect flux on the mid-plane is

$$\phi_i = -\frac{\partial V}{\partial y} \iint dS = -B_0 D = -\frac{V_0}{h_0} D. \tag{6}$$

In general, indirect flux to nearby zero scalar potential is written as:

$$\phi_i = fS' \frac{V_0}{h}. \tag{7}$$

S' is the surface area of the iron pole, h is the distance between the iron pole and the zero potential. Consider the corner effect, the scale factor $f>1$,

## 2.3 Total magnetic flux

From $\oiint B \cdot dS = 0$, total magnetic flux around the float iron pole is zero, that is the direct flux deposits on iron pole equals to the indirect flux leaves away from the iron pole, which can be expressed as:

$$\phi_d + \phi_i = 0. \tag{8}$$

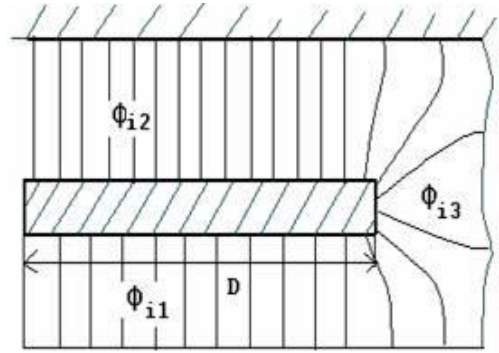

Fig. 2. The indirect flux calculation model. Indirect flux goes from the iron pole to the nearby zero scalar potential area.

### 3. H type hybrid permanent dipole design

A hybrid permanent dipole was fabricated for magnetic material processing, aimed to produce the field higher than 2.4 T in a 7 mm gap. Assume the half gap as $h_0$, the expected magnetic field at the central plane is $B_0$, then the scalar potential on the surface of the iron pole is $V_0=B_0 h_0$. Fig. 3 shows the cross section of the upper half magnet, the iron yokes are displayed in hatch. The top permanent block is magnetized along the negative y, whereas the side permanent block is radially magnetized inward. For the lower half magnet, the side permanent block is radially magnetized outward.

The direct fluxes come from the top magnetized block and the side radially magnetized permanent ring, which can be written as:

$$\phi_d = \pi R_2^2 B_r + 2\pi R_2 B_r (h_3 - h_1) \ . \tag{9}$$

The indirect fluxes scatter from the iron pole to the nearby zero scalar potential faces, which includes parts to the mid-plane, to the top yoke, to the side yoke and to the upper and lower corners. Here selects factor *f* as 1.9 to contain the corner effects. Then total indirect flux is

$$\phi_i = B_0 h_0 \left(1.9 \frac{\pi R_2^2}{h_0} + 1.9 \frac{\pi R_2^2}{h_4 - h_2} + \frac{2\pi R_2 (h_3 - h_0)}{R_3 - R_2}\right) \ . \tag{10}$$

NdFeB N44H material is selected for permanent blocks, the remnant field $B_r=1.36$ T. Other related parameters are: half magnet gap $h_0=3.5$ mm, pole tip length $h_1=10$ mm, distance from pole top to mid-plane $h_2=20$ mm, distance between the top of the side permanent blocks to the mid-plane $h_3=30$ mm, side yoke height $h_4=40$ mm, pole radius $R_2=14$ mm, pole tip radius $R_1=5$ mm, radius of the magnet gap $R_3=32$ mm, return yoke thickness 8 mm. Vanadium Iron is select as the magnetic pole, since it has high saturated field as $B_s=2.2$ T, the return yokes are made of DT4 soft iron. By eq. 10, magnetic field produced in the mid-plane can reach 2.42 T. OPERA-3d [13] software is used to check the field strength, the calculated peak field on the mid-plane is 2.45 T.

In comparison, three cases were calculated when the top permanent blocks removed or replaced with soft iron. Fig 4 shows field differences along the central mid-plane. When the top permanent block was removed, total direct fluxes were reduced, field strength on the mid-plane will drop accordingly. What's more, when the top permanent blocks are replaced with iron, part of magnetic flux will directly return the top yoke, the field in the magnet gap will decrease much more.

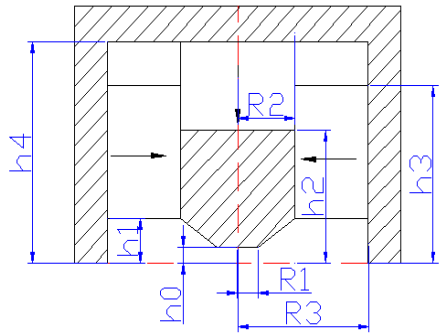

Fig 3. Cross section of the upper half hybrid permanent dipole. The material of the magnet pole and the side yoke are Vanadium and DT4 iron respectively.

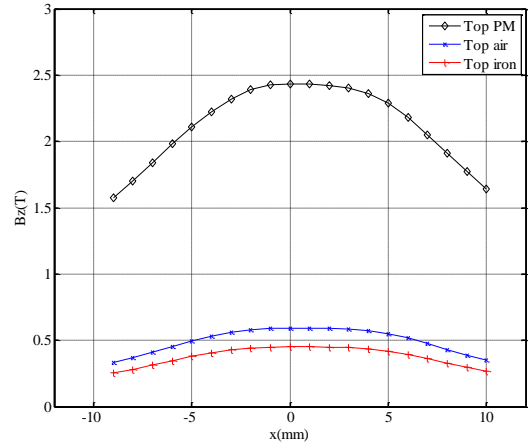

Fig 4. Field differences on the mid-plane when the top permanent magnet replaced with air or DT4 iron.

## 4. Magnet fabrication and field measurement

For technical limitation, the radial magnetized block was replaced by 6 tile-liked blocks. Fig. 5 shows the lower half magnet assebly. In order to protect the permanent blocks, they are covered by a G10 board. Fig. 6 shows the whole magnet assembly.

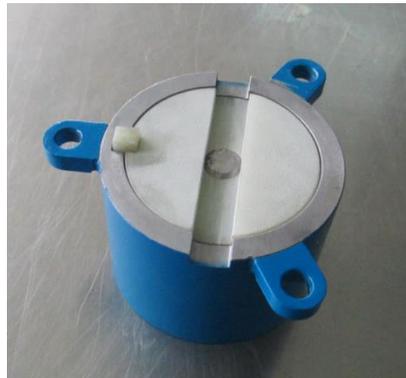

Fig.5. Lower half of the hybrid permanent magnet assembly

Field measurement was done by a Hall probe along the slot in the G10 board, Fig. 7 shows the field measurement result, it has a little difference compare with the 3D field calculation.

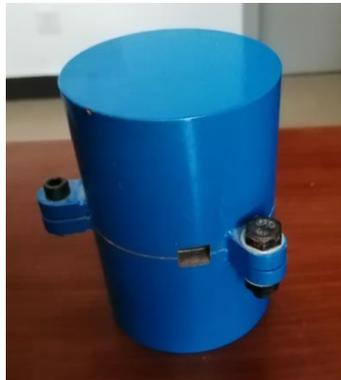

Fig. 6. Whole Hybrid permanent magnet assembly

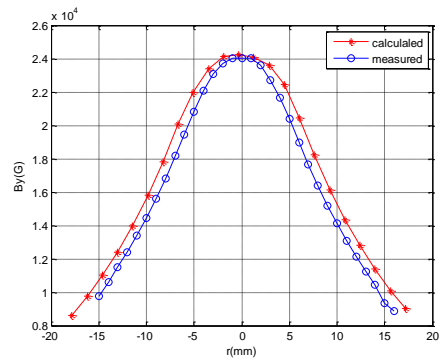

Fig. 7. Calculated and measured field distribution along the central line in the magnet mid-plane

## 5. Design variable gradient permanent quadrupole by two nested permanent rings

Variable gradient quadrupole can be built with pure or hybrid permanent magnets. Fig 8 shows a kind of pure permanent magnet design, where the variable gradient was realized by the relative rotation between the inner and outer permanent rings, field gradient varies from $G_1-G_2$ to $G_1+G_2$, here $G_1$ and $G_2$ are field gradient of the inner and outer permanent rings respectively. The nested pure permanent rings have two disadvantages, they are the lower efficiency and the accompanied skew quadrupole components.

First, permanent blocks in outer rings are much away from the inner, its field contribution are greatly reduced, which needs larger size and the cost will increase accordingly. On the other hand, since permanent blocks is similar as air, a skew quadrupole component will produce during rotation and cannot be canceled. Skew quadrupole component will give rise to work point drift, increase the beam emittance and will eventually affect the beam life time. Same problem exists in reference [15] for two sets of nest permanent rings that made of cylindrical permanent rods.

Another plan is using the hybrid permanent quadrupole, where several permanent blocks are replaced by iron poles, by which to control the field quality and concentrate the magnetic flux. Fig. 9 shows a hybrid permanent quadrupole that consists of two sets of permanent rings, the variable gradient is realized by the relative rotation between the inner and outer ring.

In a circular particle accelerator, the ramping period is in a few seconds, gradient changes for a quadrupole can go along with that of the beam energy. According to the design idea for the conventional electromagnet, the inner surface of the iron pole in a hybrid permanent magnet is selected as a part of hyperbola to increase the field uniformity [14]. The outer circular surface of the iron pole is selected as wide enough to collect the magnetic flux from the outer ring. Relative rotation between the two permanent rings does not bring extra skew quadrupole components, since the stray field is blocked by the iron pole.

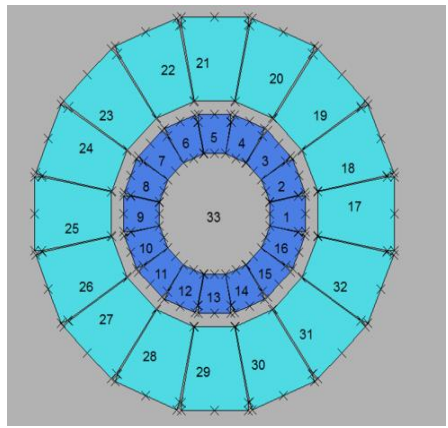

Fig 8. A variable gradient quadrupole that consists of two pure permanent rings

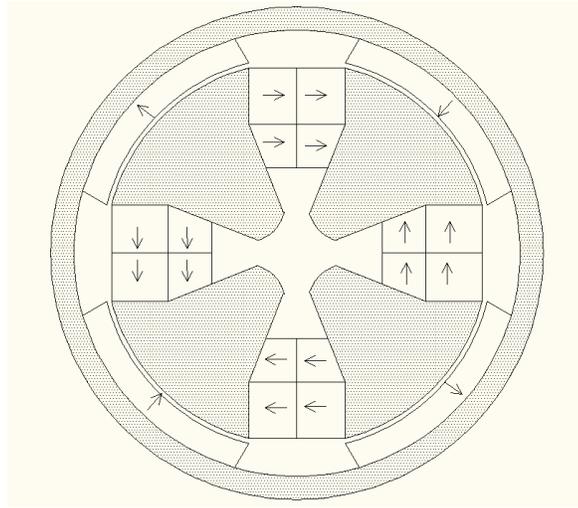

Fig. 9. Variable gradient quadrupole consists of two nest hybrid permanent rings，the dash regions are made of iron.

Fig. 10 shows the 3D field calculation when the outer ring rotated at 60 degrees, The design parameters are: magnet aperture 40 mm, outer diameter 320 mm, magnet length is 100 mm. Taking the suitable shimmed on the iron pole surface and end plate, all the high order harmonics can be reduced less than 5 units at different rotation angle. Using FFT function in OPERA-3d, the quadrupole gradient and field harmonics at the reference radius of 13 mm can be found. The calculated gradient varies from 21 T/m to 64 T/m in a 90 degrees rotation period, maximum torque is 240 N.m, which can be realized by motors with the reduction gearbox.

Fig. 12 shows the normalized high order harmonics along the beam line, all the integral harmonics are less than 5 units. Table 1 shows the normal and skew quadrupole values at different rotation angles, where all skew quadrupoles is near to zero.

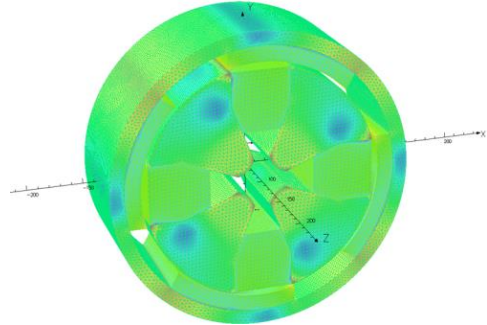

Fig. 10. Calculation example of a variable gradient quadrupole that consists of two nest hybrid permanent rings when the outer ring rotation at 60 degrees.

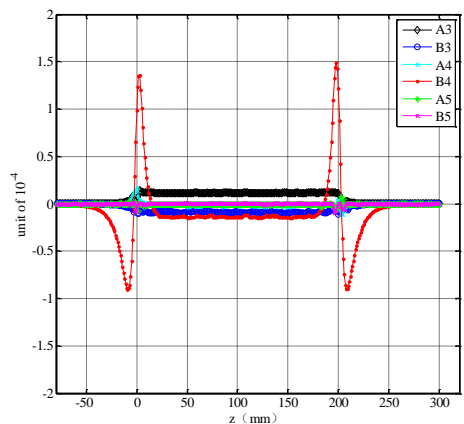

Fig 11　High order harmonics along the beam line (@r=13mm ) 　when the outer ring rotaed at 60 degrees, all data are normalized with the integral quadrupole strength.

Table 1 Normal and skew quadrupole components changes at difference rotation angles

| Rotation angles | 0 | 30 | 60 | 90 |
|---|---|---|---|---|
| B2(T/m) | 64.21 | 55.12 | 35.07 | 25.47 |
| A2(T/m) | 2.0E-004 | -0.0060 | 0.016 | 0.027 |

## 6. Design variable gradient sextupole for small-angle neutron scattering detector

A variable gradient hybrid permanent sextupole was designed for the Very Small-angle Neutron Scattering instrument (VSANS) in the China Spallation Neutron Source Science (CSNS). As shown in Fig. 12, inner permanent ring has 12 permanent blocks and 6 Vanadium Iron poles to collect the magnetic fluxes from the inner and outer permanent rings. Magnetization angle for each permanent block is 60 degrees relative to its central symmetrical axis, which can contribute 20% field strength compare with the 90 degrees from the calculation. For the 12 blocks outer ring, easy axis orientation for each permanent block is parallel or perpendicular to its central symmetrical axis. The calculated gradient varies from 7188 T/m$^2$ to19968T/m$^2$ in a rotation recyle from 0 to 60 degrees. Fig. 13 shows the 3D simulation field when the outer ring rotated at 60 degrees relative to the inner ring.

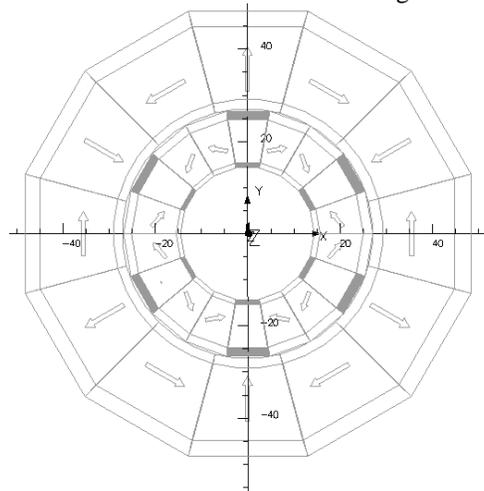

Fig. 12 , Schematic layout for the magnetic angles for inner and outer permanent ring when the outer ring at 0 degrees. The dashed areas are the iron poles.

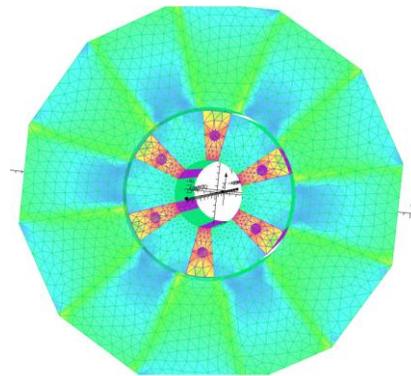

Fig 13. 3D simulated magnetic filed when the outer permanent ring rotated at 60 degrees.

For the machnical design, the inner ring is fixed on the support seat by the connected flanges at both ends, the outer ring rotates relatve to the inner ring by a set of high speed motors with reduction gearboxes. Each iron pole is a set of 5 mm sliced lamated Vanadium Irons with water cooling wholes to get rid of eddy current overheating during the 1.5 kHz high speed rotation. From 3d calculation, maximum torque is 220 N.m at 45 degrees rotation for the 200 mm long nested permanent sextupole prototype. The 200m long variable gradient sextupole has been fabricated and tested successfully.

## 7 . Conclusion

In a symmetrical hybrid permanent magnet, the mid-plane can be treated as the reference zero scalar potential, whereas the iron pole is looked as high scalar potential to collect the magnetic flux and release to the low potential area. This paper presents how it possible to produce the expected field by using magnetic fluxes method for hybrid permanent magnet design. Through theoretical calculation and 3D field simulation, a permanent dipole with field strength higher than 2.4 T was fabricated and tested. Variable gradient nested permanent quadrupole or sextupole can also be realized by using iron poles to collect the magnetic flux and block the high order harmonics from the outer ring. For its small, compact and low operation cost, hybrid permanent magnet can find more applications in areas such as particle accelerator, motor, medical equipment and material research.

99